\documentclass[11pt]{article}
\usepackage{moriond,epsfig}

\usepackage{graphicx}
\usepackage{dcolumn}
\usepackage{bm}
\usepackage{textcomp}
\usepackage{aas_macros}
\usepackage{amsmath, amssymb}

\def\be{\begin{equation}}
\def\ee{\end{equation}}
\def\bea{\begin{eqnarray}}
\def\eea{\end{eqnarray}}

\newcommand{\keVee}{~{\rm keV_{ee}}}

\newcommand{\CPY}{{\texttt{CPY}}}

\begin{document}
\title{On the Cosmic Ray Muon Hypothesis for DAMA}

\author{ Josef Pradler \thanks{jpradler@perimeterinstitute.ca}}

\address{Perimeter Institute for Theoretical Physics, 31 Caroline St N, Waterloo ON N2L 2Y5, Canada}

\maketitle\abstracts{ 
  The DAMA dark matter search experiment observes a statistically
  significant percent-level variation of its low-energy count rate
  with a period of one year.
  In this note we recall some of the arguments which challenge the
  hypothesis that the cosmic ray induced underground muon flux can be
  the origin of the modulation.
  In addition, we provide new comments on recent works on this
  subject.
}

\section{Introduction}

In a recent publication~\cite{Chang:2011eb} henceforth referred to as
\CPY\ we examine recent claims that the cosmic ray muon flux can be
responsible for generating the modulation signals seen by
the DAMA~\cite{Bernabei:2008yi,Bernabei:2010mq} and, more recently, by the
CoGeNT~\cite{Aalseth:2011wp} Dark Matter (DM) direct detection
experiments. By examining the time series of the reported DAMA and
CoGeNT signals we find the data sets differ in phase, likely in
amplitude, and potentially in their power spectrum from measurements
of the underground muon flux. A correlation analysis in \CPY\ reveals
that the data sets are incompatible and that the muon flux as the sole
source of the reported signals is excluded at high confidence.

In this note we restrict the discussion of the muon hypothesis to DAMA
which is the only direct detection experiment which claims detection
of a firm DM signal. It is situated at Gran Sasso, Italy, in the
underground LNGS laboratory. The collected data encompasses two major
runs, DAMA/NaI (Dec 1995 - July 2002) and DAMA/LIBRA (Sept 2003 - Sept
2009). The residual count rate reported by the collaboration exhibits
annual modulation compatible with what is expected from generic DM
models.

\begin{figure}[tb]
\begin{center}
\includegraphics[width=0.7\textwidth]{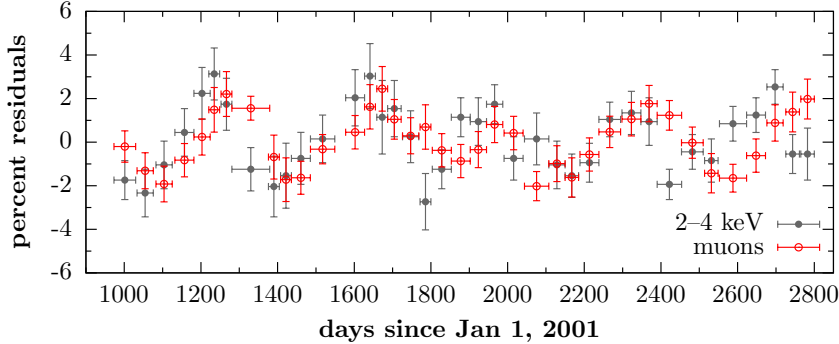}
\end{center}
\caption{Percent annual residuals of the LVD measured muon flux when
  binned in accordance with the DAMA/LIBRA runs~1--5. The latter
  residuals are shown for the 2--4\,keV bins assuming a baseline count
  rate of $\bar s = 1.15$~cpd/kg/keV.}
\label{fig:lngs}
\end{figure}

TeV-scale cosmic ray muons produced at stratospheric altitude levels
can reach deep underground and induce spallation reactions in the
detector and nearby.  The seasonal variations of the underground muon
flux have been observed on the northern and southern hemisphere
alike. For DAMA, the relevant measurement at the LNGS site is the one
from LVD~\cite{Selvi} which overlaps in time with DAMA/LIBRA cycles
1--5. The integral muon intensity underground reads $\langle I_{\mu}
\rangle \simeq 3\times 10^{-4}\,\mathrm{m}^{-2}\,\sec^{-1}$ and
exhibits an annual variation of $\sim 2\%$ in amplitude. In
Fig.~\ref{fig:lngs} we show the percent residuals of the muon flux
when binned in concordance with DAMA with the annual mean count rate
subtracted. In addition, the (2--4)\,keVee\ bin of the DAMA/LIBRA
experiment are shown; a baseline rate of $\bar s = 1.15$~cpd/kg/keV\
has been assumed. The seeming similarity in time and amplitude is
tantalizing and has lead to the notion that both signals may in fact
be measurements of the very same cosmic ray
phenomenon~\cite{ralston,nygren,Blum:2011jf}.  In the following we
recap and further develop on the arguments which challenge this
hypothesis.

\section{The Null Hypothesis: periods, binning, and subtraction of means}
\label{sec:null}

A spectral analysis will most readily reveal periodic behavior in a
time series $\{t_i, X_i\}_{i=1,\dots N}$. In case the times $t_i$ are
unevenly spaced, it is most convenient to make use of the Lomb-Scargle
(LS) periodogram~\cite{1982ApJ...263..835S},
\begin{eqnarray}
  \label{eq:LS}
  P_X(\omega) = 
\frac{1}{2 \sigma_X^2} \left\{ \frac{\left[ \sum_i (X_i - \bar X) \cos\omega (t_i-\tau) \right]^2}{ \sum_i \cos^2 \omega (t_i-\tau)} + 
 \frac{\left[ \sum_i (X_i - \bar X) \sin\omega (t_i-\tau) \right]^2}{ \sum_i \sin^2 \omega (t_i-\tau)} \right\} . 
\end{eqnarray}
Here $\tau $ is obtained from 
$  \tan(2\omega \tau) = {\sum_i \sin 2\omega t_i}/{\sum_i \cos 2\omega t_i }$,
and it ensures that $P_X(\omega)$ is invariant to shifts in the time
origin; $\bar X$ and $\sigma^2_X$ are the arithmetic mean and the
variance of the data $\{ X_i \}$, respectively.  A Bayesian derivation
and generalization of (\ref{eq:LS}) can be found in \CPY.  Noise is
rejected at the $1-\alpha$ confidence level by demanding that the
power at a sampled frequency is in excess of
\begin{eqnarray}
  \label{eq:sig}
  z_\alpha =-\ln \left[ 1-(1-\alpha)^{1/M} \right] .
\end{eqnarray}
A statistical penalty has been included for examining $M$ independent
frequencies.  In order to obtain a meaningful statistical
interpretation we have to choose $M$ carefully. In \CPY\ we chose to
sample the data at frequencies $\omega_n = n \omega_F$ with
$n=1,\dots, N$, $M=N$, and with the fundamental frequency $\omega_F =
2\pi/T$ where $T$ is roughly the range of years covered by the data
set. In this note we instead oversample the data using a finer grid of
frequencies.  Since $X_i$ are not grossly unevenly spaced in $t_i$,
$M\approx N $ holds~\cite{Horne:1986bi} as long as we only sample
$X_i$ above the average Nyquist frequency, $\nu_c = 1/2\Delta$ where
$\Delta$ denotes the average spacing between $t_i$. We checked that
both approaches agree on coinciding frequencies.

\begin{figure}[t]
\begin{center}
\includegraphics[width=0.47\textwidth]{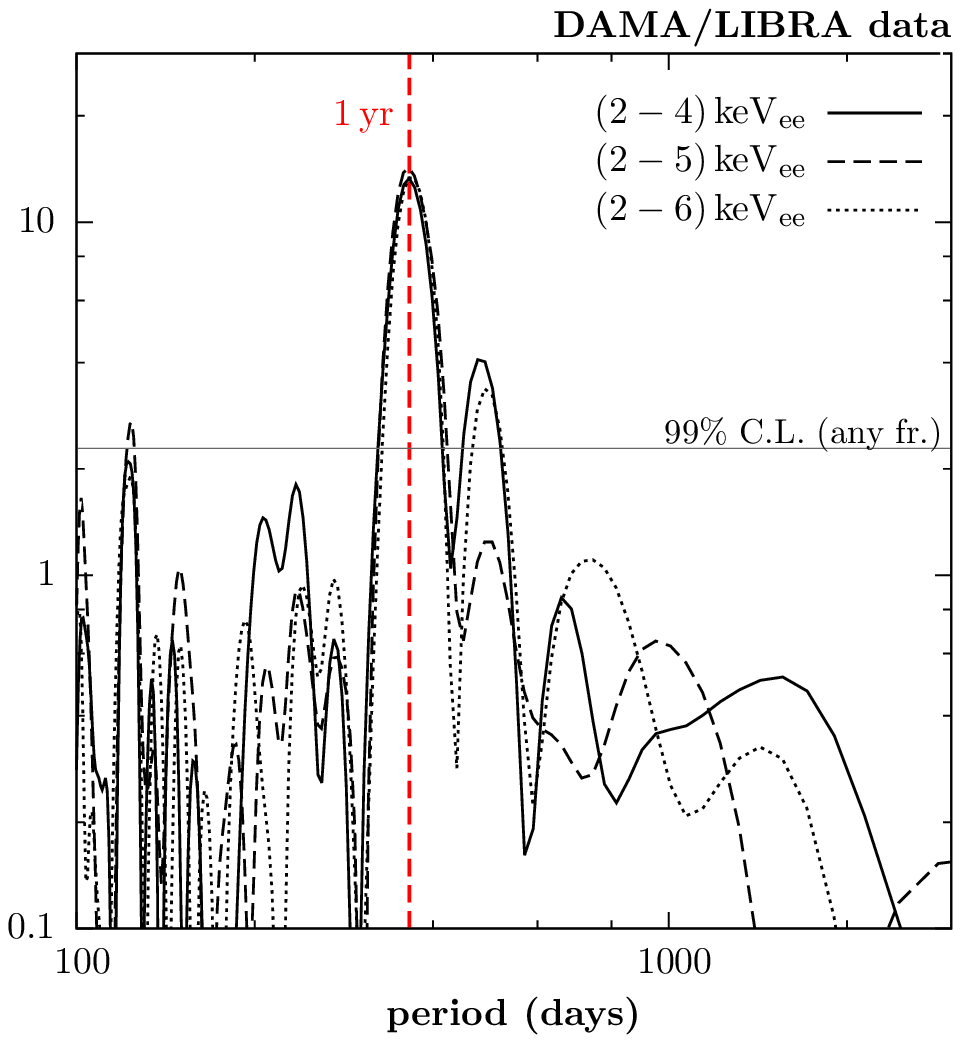}%
\hfill
\includegraphics[width=0.47\textwidth]{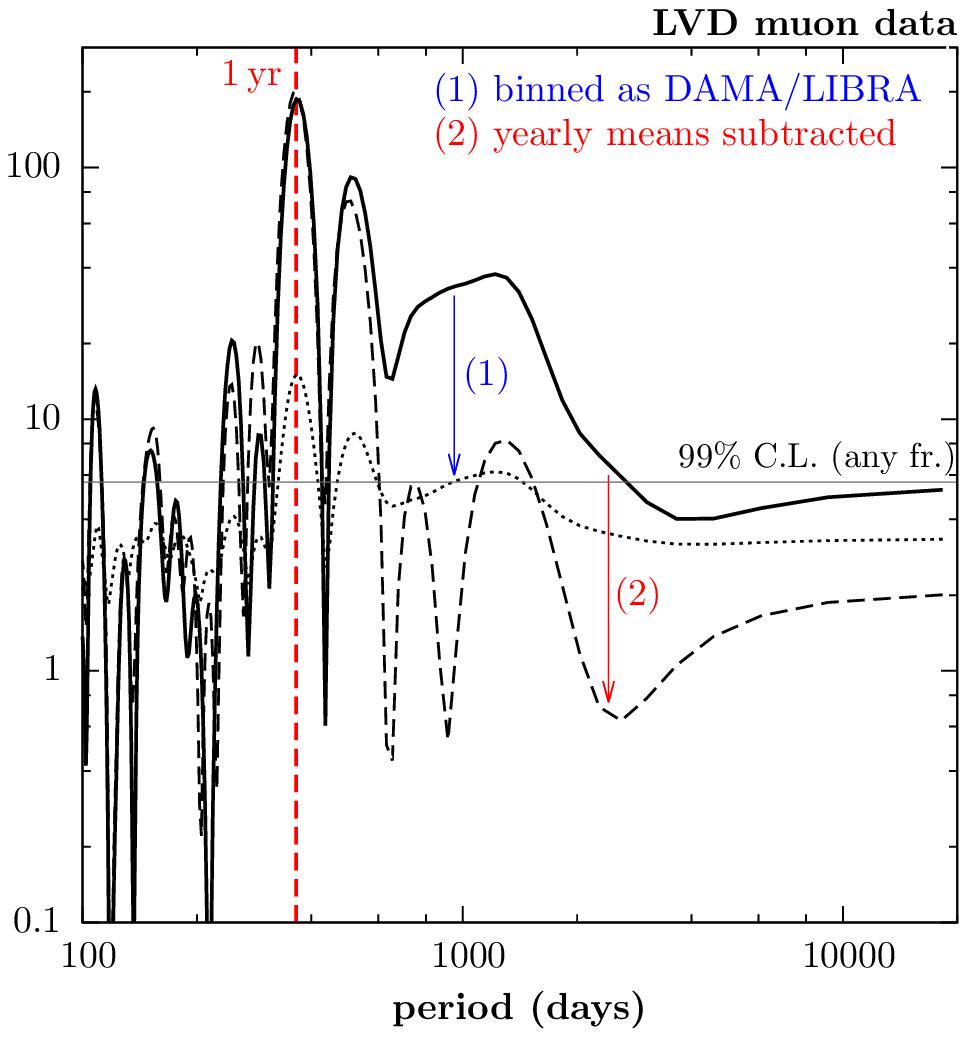}
\end{center}
\caption{ \textit{Left:} LS spectrum for DAMA/LIBRA residual count
  rates as a function of the period.  The horizontal gray line marks
  the power above which noise can be rejected with $99\%$ confidence
  for any frequency. \textit{Right:} Power spectra of the LVD data
  which has temporal overlap with DAMA/LIBRA (cycles 1--5). The solid
  line is for unbinned LVD data; the dotted line is obtained after
  binning the data in concordance with the time bins of DAMA/LIBRA
  \textit{and} after re-scaling the resulting power spectrum to match
  significance levels of the solid curve: the dotted line is no LS but
  can directly be compared with the other lines. The dashed line is
  obtained from unbinned LVD data but the yearly mean intensity has
  been subtracted. The subtraction results in damping of the long
  period modes; binning damps power on all time-scales. Both effects
  make the muon data appear more similar to DAMA.}
\label{fig:damalvd}
\end{figure}

The left panel of Fig.~\ref{fig:damalvd} shows the DAMA/LIBRA power
spectrum for the residual count rates as reported
in~\cite{Bernabei:2010mq} and which are obtained by subtracting the
mean count rate in each cycle~\cite{belli:2012} for the various energy
bins as labeled. Above the horizontal line, noise is rejected at
$99\%$~C.L.  With an approximate power $P_X(2\pi/1\,\mathrm{yr}) \sim 12$
at a period of $1$~yr, the null hypothesis is disfavored by at least
$4.5\sigma$ at a single frequency and at $3.6\sigma$ according to
(\ref{eq:sig}). Notably, there is significant power at a period at
$480\,$days in the $(2-4)\,\keVee$ and $(2-6)\,\keVee$ bins and a
consistent power in all three energy bins with period~1/3\,yr
corresponding to a triannual mode; for the latter only the
$(2-5)\,\keVee$ bin exceeds the $99\%$~C.L.~level.

The right panel of Fig.~\ref{fig:damalvd} shows the LS periodogram
obtained from LVD data where only those data points were included
which have actual temporal overlap with the bins of the DAMA/LIBRA
cycles, namely, full cycles 1--5.
We only show a fraction of the full power spectrum with smaller
periods omitted. The horizontal line is calculated according to
(\ref{eq:sig}) from all sampled independent frequencies. Since
$N_{\mathrm{LVD}}\gg N_{\mathrm{DAMA}}$ the required power to reject
the null hypothesis is larger when compared to the right plot
---despite the similar period intervals shown.
The solid line is obtained from the unbinned LVD data set. Aside from
the significant yearly modulation, the spectrum exhibits temporal
variations on time-scales smaller \textit{and} greater than one
year. At first sight this is in stark contrast to the DAMA spectra
which barely exhibit power at periods other than at
$T=1\,$yr. However, in order to better compare these data sets one
should (1) bin the muon data over the same time spans DAMA/LIBRA bins
their signal and (2) subtract the yearly mean of muon flux. The latter
mimics the procedure employed by the DAMA collaboration when they
present their residual count rates. To isolate the effects of (1) and
(2) we carry out each procedure separately and show its effect on the
power spectrum.

We first observe that binning reduces the number of data points from
$N\sim 1.3$k to $N=38$ when considering DAMA/LIBRA cycles 1--5. For a
given significance $\alpha$ this changes the condition on the minimum
power in Eq.~(\ref{eq:sig}). Therefore, we first bin the LVD data and
subsequently scale the resulting power spectrum in such a way that for
a given $y$-value the significance agrees with the one from the
unbinned data. The result is shown by the dotted line in
Fig.~\ref{fig:damalvd}. Though the dotted line is not a LS anymore,
one can now directly compare the dotted with solid line.  We observe
that the generic effect of binning is to mask power on all time
scales.
The effect of subtracting the mean intensity from the data on a yearly
basis is shown by the dashed line instead.  This time the power
spectrum is not scaled and the dashed line is indeed a LS
periodogram. In contrast to binning, this procedure largely preserves
power on periods $T\leq 1\,$yr. However, for larger periods we again
observe that variations are masked due to the particular treatment of
the data.

In contrast to the LVD spectrum, the Fourier transformed DAMA
residuals in the left panel of Fig.~\ref{fig:damalvd} show much less
significant power at periods smaller and greater than a year. As shown
above, this may simply be an artifact of the way the DAMA/LIBRA
modulation signal is obtained. It is unfortunate since a Fourier
analysis could otherwise be used to discriminate between background
induced effects and real signal.  It is not clear how to assign a
quantitative significance to this difference without knowledge of the
full, unsubtracted time series. This calls for a release of the yearly
baseline count rates by the DAMA collaboration.

The DAMA collaboration has recently responded to the criticism that
baselines are subtracted per cycle~\cite{Bernabei:2012wp}. Instead of
providing the actual values, the LS power spectrum of the joint
DAMA/NaI and DAMA/LIBRA baselines is presented. No significant power
which would point towards long-term behavior is observed. However, the
baselines are a mere set of 13~numbers. It is not surprising that no
statistically significant power is observed from such a small data
set. Similarly, we find that the LVD baseline spectrum does not yield
any significant power when calculated according to~(\ref{eq:LS}), too.
This casts doubt that a presentation of the baseline-LS precludes the
existence of variations in DAMA with periods larger than one year,
similarly to what is observed in muons; see
also~\cite{UCLAjosef,UCLAkfir}.

\section{The phase of muons: a potentially stretched concept}
\label{sec:phase}

Given an isotropic DM velocity distribution $f(v)$ in the earth's
vicinity, the differential recoil rate $dR/dE_R$ of DM scattering in
the detector is predicted to peak on June 2$^{\rm nd}$, corresponding
to a phase of $t_0 =152.5$ days after January 1$^{\rm st}$,
\begin{eqnarray}
  \label{eq:drder}
  \frac{dR}{dE_R}& \propto &\int^\infty_{v_{\mathrm{min}}} \frac{f(v)}{v}dv
  \approx
  c_0 + c_1 \cos{\left[\omega(t-t_0)\right]} .
\end{eqnarray}
Here, $v_{\mathrm{min}}$ is the minimum required relative velocity
between DM and target which produces a nuclear recoil of energy
$E_R$. In the canonical case, the annual mode is at the percent level,
$c_1/c_0 = \mathcal{O}(10^{-2})$; higher harmonic corrections to
(\ref{eq:drder}) have first been pointed out in~\CPY.
The clear-cut prediction on $t_0$ arises from the geometric setup of the
earth's rotation around the sun in conjunction with the movement of
the solar system relative to the Galactic coordinate system. Hence,
apart from the period, the second most important characteristic of the
oscillations observed by the DAMA experiment is the phase of the
signal. 

On the other hand, when considering environmental factors as potential
background, the variation in time will in general not be modulated
with a strict period of a year. Additionally, we cannot expect a
sinusoidal behavior similar to (\ref{eq:drder}). Both assertions can
be tested on the LVD and DAMA data sets. For this we model the data by
a sinusoid, minimize the usual $\chi^2$ function, and assess the
goodness-of-fit. Confidence regions in $t_0$ and $T$ can be inferred
from a $\Delta \chi^2$-method which is equivalent to constructing a
profile likelihood.
Fixing the period to one year we find that the DAMA/LIBRA
$(2-4)\,\keVee$ and the full LVD data do not agree with respective values
\begin{align}
  \label{eq:phase-posterior}
\mathrm{DAMA/LIBRA:}  & \quad t_0  = (130\pm 8)\,\mathrm{days}, \quad \chi^2/dof = 37.5/41 ,  \\ 
\mathrm{full\ LVD:} & \quad t_0  = (185\pm 1.5)~\mathrm{days}  , \quad \chi^2/dof =  3450/1971 .
\end{align}
The uncertainty quoted is statistical. An additional uncertainty of
$\pm 2$~days may be attributed to digitization. A few comments are in
order: 1) The sinusoid provides a very poor description of the LVD
data with prohibitively large $\chi^2$. This supports our expectation
that a complex phenomenon such as the underground muon flux cannot be
adequately described by this simple model. 2) Even if we scale up the
error bars for LVD to the point where $\chi^2/dof\simeq 1$, the
stringent error on the phase remains below $\pm 2$~days. 3) An error of $\pm
15$~days on $t_0$ as quoted by LVD~\cite{Selvi} is certainly
incorrect. 4) Once $T$ is allowed to float, the confidence regions in
$t_0$ and $T$ become sensitive to the time-origin; for further details
we refer the reader to \CPY. These results have recently been
confirmed in~\cite{FernandezMartinez:2012wd}. One major shortcoming of
this approach is that the data has to be subjected to the simple model
of sinusoidal variation, before any conclusions can be drawn. Indeed,
the power-spectrum of the LVD data makes it clear that significant
power exists in modes with periods larger and smaller than one
year. The notion of a single phase must therefore be treated with
caution.

\section{A correlation analysis: the conclusive approach}
\label{sec:correlation}

A glance at Fig.~\ref{fig:lngs} suggests that the DAMA/LIBRA and LVD
data sets are correlated. Indeed, when evaluating Pearson's
coefficient of correlation, $r\in[-1,1]$, one finds for the DAMA and
LVD sets as presented in Fig.~\ref{fig:lngs} 
\begin{eqnarray}
r_{\mathrm{LVD, DAMA}} = 0.44. \label{eq:corr}
\end{eqnarray}
This value excludes the no-correlation hypothesis with a confidence
level greater than 99\%. However, just because the data sets are
correlated does not imply that they are causally connected. This can
only be answered on a model-by-model basis.

Clearly, a muon background model for DAMA must encompass the
stochastic nature of the underlying process. Even if the timing
between muons and DAMA are incommensurate at first sight, could it be
that such a Poisson smearing alleviates the observed tension in the
annual phase? That this is indeed the case has been shown
in~\cite{Blum:2011jf} where a simple, linear model for how a
muon-sourced background may be realized is employed. Taking from this
that the muon hypothesis for DAMA would become viable again is however
not correct.

In \CPY\ we generate $10^4$ realizations for DAMA based on the generic
model presented in~\cite{Blum:2011jf} which connects the muon flux
measured in LVD with the count rate observed in DAMA/LIBRA. The
question which has to be asked is now the following: How likely is it
that one realization out of the mock data indeed induces the signal
observed in DAMA. To answer this question we can again look at the
correlation coefficient. It has the merit that we do not impose any
functional form on the data. As expected, the generated mock data
exhibits a high degree of correlation with LVD. In fact, the
correlation with the actual muon data is substantially \textit{higher}
than the one which is actually observed in~(\ref{eq:corr}). This rules
out the model at $99\%$~C.L.. Given the generality of the model, it
also strongly disfavors the hypothesis that muons can be the origin
for the count rate variation seen in the DAMA detectors.

\section{Conclusions}
\label{sec:conclusions}

In \CPY\ we lay out a detailed time-series analysis of DM direct
detection data as well as related datasets. In particular, we find no
evidence to support the claim that atmospheric muons are responsible
for the signal that the DAMA collaboration observes. We identify
difficulties in phase, amplitude, power spectrum and degree of implied
correlation between the data sets. Here we offer some additional
comments:
\begin{list}{\labelitemi}{\leftmargin=1em}
\item When binning data, fine-grained information on the time-series
  is lost. In addition, we find that when the LVD measured muon flux
  is binned in time in the same way as DAMA/LIBRA presents their data,
  significant power is lost on \textit{all} time scales. This has the
  consequence that the LVD and DAMA data sets become more alike in the
  frequency domain with a dominating peak at a period of one
  year. Thus binning seems to diminish discriminating potential
  between the data sets. This calls for an unbinned analysis of the
  DAMA data set.
\item The uncertainty on the phase of the underground muon flux
  reported by LVD, $t_0 = (185 \pm 15)$ days is erroneous. In \CPY, an
  uncertainty of $ \pm 2$~days was obtained, but the discrepancy had
  not been emphasized. An independent analysis
  in~\cite{FernandezMartinez:2012wd} confirms our value.
\item A recent paper by the DAMA collaboration presents the LS spectra
  of the average count rates in the DAMA/NaI and DAMA/LIBRA cycles. No
  significant power indicating potential long-term variation
  ($T>1$~yr) is observed. Following this procedure for LVD data, we
  can reach analogous conclusions---in seeming contradiction with the
  power observed in the right panel of Fig.~\ref{fig:damalvd}. One
  cannot expect statistical significance from a LS-transform of
  $\mathcal{O}(10)$ numbers without further insight into the
  normalization of (\ref{eq:LS}).
\item The same paper confirms our previous findings in \CPY\ that a
  putative cosmic ray activation can only lead to a maximum phase
  shift of a quarter year in the subsequently modulated decay. Hence
  it is not possible to overcome the $\sim 11$ month discrepancy
  between muons and DAMA.
\item A recent paper~\cite{FernandezMartinez:2012wd} incorporates
  additional data on the Gran Sasso underground muon flux. Frequentist
  fits to period and phase improve on the already significantly
  disjoint LVD and DAMA regions presented in \CPY. This corroborates
  the statement that both data sets are seemingly
  incompatible. However, as argued in Sec.~\ref{sec:correlation} and
  at great length in \CPY\ reliance on the inferred phase can be
  misleading. Indeed in~\cite{Blum:2011jf} and \CPY\ it was shown that
  Poisson smearing can sufficiently alleviate the tension between the
  phases of the muon and DAMA data sets. It is the correlation
  analysis which reveals that the data sets remain nevertheless
  incompatible. However, the latter conclusion depends on the model
  which connects the muon flux with the DAMA count rate and it can
  only be achieved considering data sets which have actual temporal
  overlap. From this perspective, no further gain of information with
  respect to DAMA is obtained by enlarging the set of muon
  measurements as done in~\cite{FernandezMartinez:2012wd}.
\end{list}
As in the case of DAMA, in \CPY---employing a similar line of
analyses---we find no significant correlation between the CoGeNT data
set and the yearly variation of the muon flux. In summary, cosmic ray
muons cannot be the sole source for the observed modulation signals in
these experiments.

\section*{Acknowledgments}
It is a pleasure to thank S.~Chang and I.~Yavin for collaboration on
this subject. The author also acknowledges useful discussions with
K.~Blum and E.~Fernandez-Martinez.

\bibliography{biblio}

\begin{thebibliography}{10}

\bibitem{Chang:2011eb}
Spencer Chang, Josef Pradler, and Itay Yavin.
\newblock {Statistical Tests of Noise and Harmony in Dark Matter Modulation
  Signals}.
\newblock {\em Phys.Rev.}, D85:063505, 2012.

\bibitem{Bernabei:2008yi}
R.~Bernabei et~al.
\newblock {First results from DAMA/LIBRA and the combined results with
  DAMA/NaI}.
\newblock {\em Eur. Phys. J.}, C56:333--355, 2008.

\bibitem{Bernabei:2010mq}
R.~Bernabei et~al.
\newblock {New results from DAMA/LIBRA}.
\newblock {\em Eur.Phys.J.}, C67:39--49, 2010.

\bibitem{Aalseth:2011wp}
C.E. Aalseth, P.S. Barbeau, J.~Colaresi, J.I. Collar, J.~Diaz~Leon, et~al.
\newblock {Search for an Annual Modulation in a P-type Point Contact Germanium
  Dark Matter Detector}.
\newblock {\em Phys.Rev.Lett.}, 107:141301, 2011.

\bibitem{Selvi}
M.~Selvi.
\newblock {Analysis of the seasonal modulation of the cosmic muon flux in the
  LVD detector during 2001-2008}.
\newblock {\em Proceedings of the 31st {ICRC}}, 1, 2009.

\bibitem{ralston}
John~P. Ralston.
\newblock {One Model Explains DAMA/LIBRA, CoGENT, CDMS, and XENON}.
\newblock 2010.

\bibitem{nygren}
David Nygren.
\newblock {A testable conventional hypothesis for the DAMA-LIBRA annual
  modulation}.
\newblock 2011.

\bibitem{Blum:2011jf}
Kfir Blum.
\newblock {DAMA vs. the annually modulated muon background}.
\newblock 2011.

\bibitem{1982ApJ...263..835S}
J.~D. {Scargle}.
\newblock {Studies in astronomical time series analysis. II - Statistical
  aspects of spectral analysis of unevenly spaced data}.
\newblock {\em \apj}, 263:835--853, December 1982.

\bibitem{Horne:1986bi}
J.H. Horne and S.L. Baliunas.
\newblock {A Prescription for period analysis of unevenly sampled time series}.
\newblock {\em Astrophys.J.}, 302:757--763, 1986.

\bibitem{belli:2012}
G.~Belli.
\newblock {private communication}.

\bibitem{Bernabei:2012wp}
R.~Bernabei, P.~Belli, F.~Cappella, V.~Caracciolo, R.~Cerulli, et~al.
\newblock {No role for muons in the DAMA annual modulation results}.
\newblock 2012.

\bibitem{UCLAjosef}
J.~Pradler.
\newblock {Talk given at the UCLA Dark Matter meeting, Marina del Ray, 2012.}

\bibitem{UCLAkfir}
K.~Blum.
\newblock {Talk given at the UCLA Dark Matter meeting, Marina del Ray, 2012.}

\bibitem{FernandezMartinez:2012wd}
Enrique Fernandez-Martinez and Rakhi Mahbubani.
\newblock {The Gran Sasso muon puzzle}.
\newblock 2012.

\end{thebibliography}
\end{document}